\documentclass{sig-alternate}
\usepackage[pdftex]{color}

\usepackage{cite}
\usepackage{graphicx}
\usepackage{url}
\usepackage{balance}

\newcommand{\eg}{e.\,g.\ }

\begin{document}
%
\conferenceinfo{ROA'08,} {May 11, 2008, Leipzig, Germany.}  
\CopyrightYear{2008} 
\crdata{978-1-60558-028-7/08/05}

\title{Abstractness, Specificity, and Complexity\\in Software Design}

\numberofauthors{1}

\author{
  \alignauthor Stefan Wagner and Florian Deissenboeck\\
    \affaddr{Institut f\"ur Informatik}\\
    \affaddr{Technische Universit\"at M\"unchen}\\
    \affaddr{Garching b.\ M\"unchen, Germany}
}

\maketitle

\begin{abstract}

Abstraction is one of the fundamental concepts of software design. 
Consequently, the determination of an appropriate abstraction level for the 
multitude of artefacts that form a software system is an integral part of 
software engineering. However, the very nature of abstraction in software 
design and particularly its interrelation with equally important concepts 
like complexity, specificity or genericity are not fully understood today. 
As a step towards a better understanding of the trade-offs involved, this 
paper proposes a distinction of abstraction into two types that have 
different effects on the specificity and the complexity of artefacts. We 
discuss the roles of the two types of abstraction in software design and 
explain the interrelations between abstractness, specificity, and 
complexity. Furthermore, we illustrate the benefit of the proposed 
distinction with multiple examples and describe consequences of our 
findings for software design activities. 

\end{abstract}




\vspace{1em}
\begin{quote}
\emph{
``In the development of the understanding of complex phenomena, the most
powerful tool available to the human intellect is abstraction.''}

\begin{flushright}
{\small Sir Tony Hoare\cite{hoare72}}
\end{flushright}
\end{quote}

\newpage

\section{Introduction}

Abstraction is an important tool in any engineering discipline but in 
software engineering it is essential. Everything we build during the 
development of software systems contains abstractions. Textual 
requirements, a high-level architecture view, low-level design and 
even actual code are abstractions from the real world and from the 
machine the software will be executed on. In all steps of software 
development, we are confronted with decisions about what abstractions to 
use. In particular, we often have to decide on what level of abstractness 
different artefacts should reside. The motives for abstraction are manifold 
but most commonly improved reuseability or reduced complexity are the 
goal.

For example, the design of a function that retrieves specific data from a 
database raises a number of questions: Will similar functionality be needed 
in other parts of the software? Hence, should the function abstract from a 
specific type of data? If it does, will it require additional parameters to 
support different types of data? It is well known that it is not always 
advisable to make everything generic and parameterisable
(``over-engineering'' or ``gold-plating'') as the costs may outweigh the benefits 
due to increased complexity. On the other hand we do not only abstract to 
allow reuse but to facilitate understandability by reducing complexity. 
Some well-chosen abstractions help any reader of an artefact -- even its 
author -- to comprehend the design and implementation more quickly 
\cite{2006_deissenboeckf_naming, 
1993_biggerstafft_comprehension_concept_assignment, 
1995_mayrhausera_comprehension_maintenance}. Still, it is not always best 
to have the highest level of abstraction possible. It is a common saying 
that we cannot understand something if it is \emph{too} abstract. These 
questions illustrate that choosing the \emph{right} abstraction level is 
non-trivial and involves complex trade-offs.

We use ``artefact'' as the basic unit of our consideration of abstraction 
in software design. An artefact can be anything that is created for 
developing software. However, for our discussion of abstractness only 
human-readable documents are of interest. Hence, an artefact can be a 
function, class, procedure, module, component, etc. All of these have a 
level of abstractness that needs to be decided on during design.

\subsection{Problem}
Despite the uttermost importance of abstraction in software development,
the implications and trade-offs involved are not totally understood. What
does it mean when I replace one or more artefacts by a more abstract one? What
implications does it have on the complexity? Are there other properties of the
artefacts involved? To our knowledge, there is no established basis to answer
these questions today although they influence nearly every task in software
development.

\subsection{Contribution}
As a first step to answer these questions, we propose three characteristics
of artefacts that capture the necessary properties involved:
\emph{abstractness}, \emph{specificity},
 and \emph{complexity}. 
These characteristics are of a relative nature and can only be analysed
w.r.t.\ other artefacts.
\emph{Abstractness} is the degree of information loss an artefact
has, \emph{specificity} denotes the number of contexts it can be
used in, and \emph{complexity} is divided into \emph{detail
complexity} and \emph{dynamic complexity}. The former is related to the
number of elements, the latter to cause-effect relationships. 
This division allows to describe
the effects of abstractions more precisely. Based on these characteristics,
we analysed a set of examples that lead to
several consequences for using abstractions.

%
\subsection{Outline}
In Sec.~\ref{sec:communication} we describe our basic view on software
development and reuse including the decisive characteristics.
 Based on these characteristics, we describe two types of
abstractions in Sec.~\ref{sec:abstractions}.
We illustrate the characteristics and abstraction types w.r.t.\ several existing
languages in Sec.~\ref{sec:examples}. 
We derive consequences of our findings for different
areas of software development in Sec.~\ref{sec:consequences}. Related
work is compared in Sec.~\ref{sec:related} and final conclusions are
given in Sec.~\ref{sec:conclusions}.


\section{Abstractness, Specificity, and \\Complexity}
\label{sec:communication}

This section sets the scene for the following discussion about the effects of
abstraction in software design. We first discuss the aims of abstraction
and what kind of artefacts we consider. Then we describe the two levels
involved in abstraction and what role the fixed and variable parts play.
Finally, the three decisive characteristics of artefacts with respect to
abstraction are defined: abstractness, specificity and complexity.

\subsection{Aims of Abstraction}

Abstraction is often said to have the aim of reducing complexity. However, 
we see complexity reduction as only one aspect and consider
the following two reasons to be fundamental for abstraction.
Firstly, abstraction is needed in order to be able to 
comprehend the necessary artefacts for software design. Reality and the 
software that interacts with reality is too complex to be understood as a 
whole. Hence, we need to divide it into smaller chunks and throw explicit 
information away in order to understand certain aspects. 

Secondly, as is often pointed out 
(cf.~\cite{krueger92,parnas89,standish84}), reuse is inexorably tied to 
abstraction. We need to raise the level of abstraction of an artefact in 
order to be able to use it in different contexts. An artefact that is 
concretely shaped for a specific context cannot be considered very 
abstract. Abstracting it to a more general (or generic) artefact enables 
the reuse of the artefact. This is obviously also connected to the first 
part. Only an artefact that can be comprehended with reasonable effort will 
be reused.

Both aims of abstraction are related to communication. One could see the 
combination of all abstractions in an artefact as a language that is used 
to communicate between different developers or designers. The names of the 
abstractions that are introduced constitute the words of the language and 
the composition rules of the abstractions constitute the grammar. We use 
different language elements and combine them in certain ways, hence, we 
communicate. This communication between humans -- as opposed to the 
communication with the computer -- becomes more and more important. With 
advances in computational power and compiler technology the details of 
\emph{how} the program is realised on the computer are less significant. 
Moreover, considering the sheer size of many of todays software systems and 
the time they are maintained, communication with humans via the programs is 
vital.

\subsection{Artefacts in Software Design}

When we discuss in this paper \emph{artefacts} in software design, we mean all
work products that are needed for the specification and realisation of
the software. Hence, the range goes from algebraic specifications, to
executable models, from conceptual models to compilable procedures and
classes. We consider all levels of software design because the design
decisions and effects are all of a similar nature w.r.t.\ abstractness,
specificity and complexity. Because we consider communication between humans
as the main issue, the artefacts are all human-readable.

There is one important difference when considering executability. In case 
we want to execute the artefact on a computer, there must be a mechanism in 
place that is able to generate the information that has been abstracted 
away. Otherwise, the computer will not know how to execute the artefact. 
For non-executable artefacts such as conceptual models, this is not 
necessary because the human reader is expected to be able to 
complete the information by himself. For example, a C compiler knows how
to generate machine code for details like register allocation that is
not made explicit in the source code (``abstracted away''). However, there is
no program that is capable of executing a UML use case diagram as the
information that is missing to generate an executable program cannot be
reconstructed in an automated manner.

\subsection{Two Levels of Abstraction}
\label{sec:big_picture}

\begin{figure}[h]
\begin{center}
  \includegraphics[width=0.8\linewidth]{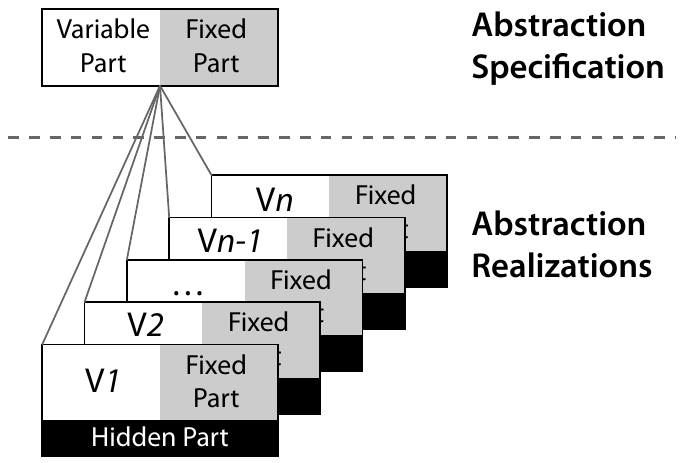}
  \caption{Specification $\rightarrow$ Realisation~\cite{krueger92}}
  \label{fig:abstraction}
\end{center}
\end{figure}

In his seminal paper on software reuse~\cite{krueger92}, Krueger presented
a simple but powerful model that illustrates the nature of abstraction
in software design. He starts from the assumption that abstraction is
a concept that cannot be discussed for a single \emph{artefact}
but needs to be discussed w.r.t. \emph{two} artefacts that are part of
an \emph{abstraction relation}. Here one of the artefacts is called
the \emph{abstraction specification} and the other is called 
\emph{abstraction realization}. The relationship is also shown in
Fig.~\ref{fig:abstraction}. In the later discussion about the effects of
abstractions we usually mean the abstraction specification. 

An abstraction specification always consists of a \emph{variable} and a 
\emph{fixed part}. The fixed part is what is set by the abstraction, 
i.e.~the information that has been abstracted but is still visible. The 
variable part is the part that can be set when realising the abstraction. 
Hence, we can have a set of abstraction realisations that ``instantiate'' 
the abstraction specification. In the realisations, there is also a hidden 
part that is added. It is also fixed but not directly visible 
from the abstraction specification.

\begin{figure}[h]
\begin{center}
  \includegraphics[width=\linewidth]{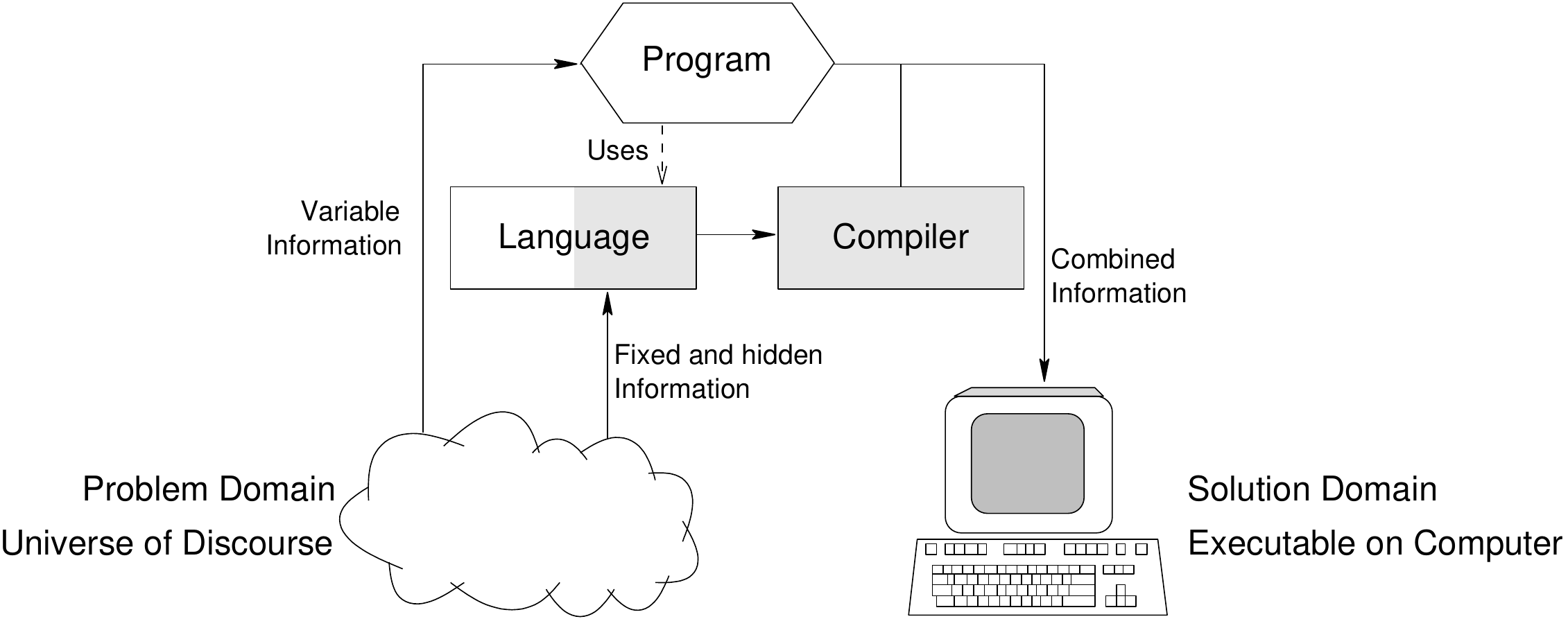}
  \caption{Variable and fixed parts in programming languages}
  \label{fig:big}
\end{center}
\end{figure}

As an example, consider this concept of two levels of abstraction for
(domain-specific) programming languages as illustrated in Fig.~\ref{fig:big}.
The language, its elements, syntax and control-structure, is the set of
abstraction specifications. It defines fixed parts that are always the same
in the language such as the paradigm used or the hardware mapping. The
variable parts are defined in the way the language elements can be combined
by the programmer. All this information is derived from the problem domain
the language aims at. A programming language used in embedded systems that
are closely tight to the hardware such as C has usually quite different 
fixed parts
than a hardware-independent language such as Java. The mapping of the objects
in Java to the memory is hidden to the programmer. In C, the programmer has
explicit means to manipulate the memory. Hence, depending on the language and
its problem domain, we have different separations into the fixed and variable
part.

The fixed and hidden parts are all encoded into the compiler of the language.
This way, they can be easily added during the compilation of a program. The
program fixes the variable parts of the language -- the abstraction
specification -- and feeds this information into the compiler. Together this
information can be put into the solution domain. Hence, an executable for a
computer system is created.

\vspace{1em}

\subsection{Characteristics}

Three important characteristics of an artefact are needed to discuss
the effects and trade-offs of abstractions: \emph{abstractness}, 
\emph{specificity},
and \emph{complexity}. We discuss each of these in the following
in more detail. Note that the characteristics are not intrinsic
properties of the artefacts but they can only be analysed in the context with
other artefacts.

\newpage
\subsubsection{Abstractness} In computer science, abstraction is always
connected to information loss \cite{guttag02}. We remove 
explicit detail and thereby
build models. Hence, abstractness, i.e., the degree of being
abstract, is determined by the amount of visible, variable information contained
in an artefact (cf.~Sec.~\ref{sec:big_picture}). 


This view fits also to \emph{word
abstractness} used in linguistics as discussed by Kammann and Streeter
\cite{kammann71}.
The abstractness of a word
is the number of subordinate words it embraces. In this definition, the
more abstract word has less information as well.
A simple example would be that ``furniture'' is more
abstract than ``chair''. The chair has more explicit information such
as that it has four legs and that it has space for a single person to
sit on. However, Kammann and Streeter discussed that
the containment relation is not always easy to define. This is also
the case in software design. 

\subsubsection{Specificity} Specificity is a characteristic that every
programmer is familiar with. It is an often occurring question how
\emph{specific} or -- in contrast -- \emph{generic} a certain solution
should be realised. In essence, this means that the specificity of an
artefact is defined by the number of contexts it can be used in. This is again
related to the observation that there are variable and fixed parts in
artefacts described in Sec.~\ref{sec:big_picture}. The larger this variable
part, the more generic is the artefact. The larger the fixed part, the more
specific is the artefact.

For example, an abstract GUI builder language probably 
has the element \emph{window} that represents the standard window in the user
interface. The specificity is then determined by the variability in the
\emph{windows}. A large variable part (e.g.~a high degree of parameterisation)
allows a use in many contexts. Hence, the specificity is low. A small
variable part causes high specificity.

\subsubsection{Complexity} The complexity of an artefact needs to be considered
because it is one of the major aims of abstraction to reduce complexity.
However, we found that there does not exist a single agreed 
definition of complexity
and even philosophy has not agreed on a unified view. Following Backlund
\cite{backlund02}, complexity is ``a measure of the effort [\ldots] that is
required to understand and cope with the system.'' This allows to measure the
complexity of the system. However, it is too coarse-grained to describe 
the effects of abstraction in software design in detail. 

In our context, the
categorisation of complexity provided by Senge \cite{senge90} is most
useful. He distinguishes \emph{detail complexity} from \emph{dynamic
complexity}. Detail complexity is what is often measured by complexity
metrics, the number of parts of an artefact. Hence, the difficulty lies in
overlooking the amount of details. Dynamic complexity, on the other hand,
describes subtle cause and effect relationships, i.e.~when it is not clear
what effects certain inputs or changes will have. In a software library,
the dynamic complexity is how hard it is to understand
how to use the interfaces and what effects changing parameters have.

\section{Types of Abstraction}
\label{sec:abstractions}

\begin{figure}[h]
\begin{center}
  \includegraphics[width=0.8\linewidth]{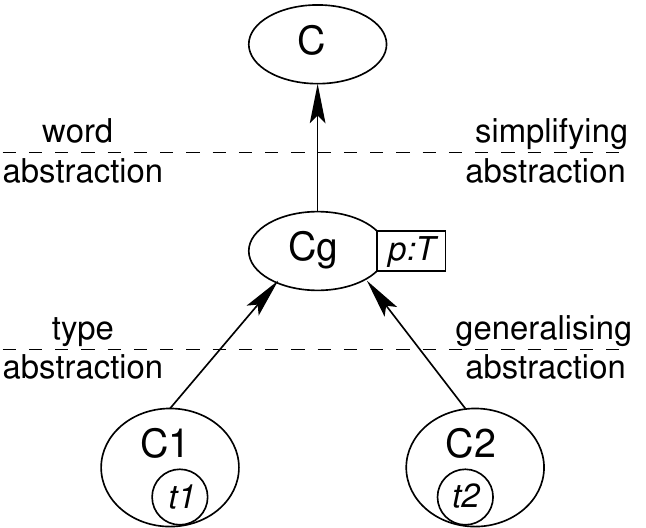}
  \caption{The two types of abstraction}
  \label{fig:abstractions}
\end{center}
\end{figure}

Having defined these characteristics, we need to analyse how abstraction in 
software design influence these characteristics. Abstraction is a central 
activity in software development. We abstract from detailed concepts to 
make them more comprehensible and to handle them all in an uniform way. 
This actually includes two different types of abstraction: (1) simplifying 
abstraction and (2) generalising abstraction. Fig.~\ref{fig:abstractions} 
illustrates these two types with related artefacts -- classes for example 
-- on two levels of abstractness.

\vspace{2em}
\subsection{Generalising Abstraction} We start with the abstraction
that is the transition from the lowest level in Fig.~\ref{fig:abstractions}
to the middle level. The \emph{type} or
\emph{generalising abstraction} is a common activity when designing
or writing software. We identify several artefacts 
that have many similarities and only differ in some
aspects. In Fig.~\ref{fig:abstractions} the differing information
between the artefacts $C_1$ and $C_2$ is only $t_1$ and $t_2$. We then 
generalise $C_1$ and $C_2$ to $C_g$ that has a parameter $p$ which
can process $t_1$, $t_2$, and anything of type $T$. Here we see that
$T$ must be some kind of super-type of the types of $t_1$ and $t_2$.
That is why we also call it \emph{type abstraction}.
However, by generalising, the type can be more than a simple union.
Using this generalising step we also become more abstract. We loose
the explicit information $t_1$ and $t_2$ and we have to input it when
using $C_g$. The major design goal for generalising abstraction
is reuse. We want to use the common aspects of $C_g$-like concepts
several times. However, we see that we did not necessarily reduce complexity
by the abstraction. 

As an example consider that $C_1$ and $C_2$ are both GUI dialog
windows that have the aim to choose an element from a tree.
For the sake of the example, the only differing information
is the title of the dialog window. In other words, $t_1$ and
$t_2$ are strings that contain the window title. In our example
this means that $t_1 = \mbox{"Choose source file"}$ and $t_2 = 
\mbox{"Choose destination file"}$. The generalising abstraction
to $C_g$ would be a dialog with a parametric title. 
As both differing
informations have the type \texttt{String}, we can simply introduce
the parameter $p$ of that type. The dialog window will then use
the value of $p$ as the title of the window. By this, we made
the language element more generic because we are now able to
use any kind of string as the window title. 

\subsection{Simplifying Abstraction} The transition from the middle
layer to the top layer in Fig.~\ref{fig:abstractions} is the
{word} or {simplifying abstraction}. This is the type of abstraction
that is used when we want to reduce dynamic complexity. In 
Fig.~\ref{fig:abstractions} we remove the information about
$p$ completely by abstracting from $C_g$ to $C$. We do not talk explicitly
about $p$ anymore. This makes the usage of $C$ simpler
-- less complex -- than the usage of $C_g$. In software design
this means that we have some fixed value $f$ for $p$
that is included in the artefact. This fixed value $f$
is strongly dependent on the context. It needs to be a superordinate
word of all the words it abstracts from. In any case, we will be
more specific. We were able to put anything into $C_g$ as long as it
is of type $T$. We are probably not able to find an abstract word
that embraces all of these possibilities and still makes sense in a realistic
application.

In the dialog window example from above, we try to remove
the parameter $p$ that takes the window title as input. This reduces
the complexity of the artefact because the artefact user
does not have to care about the window title anymore. However, this
leaves the artefact designer with the task
of finding a window title that is generic \emph{enough}. In our
case we could use the title ``Choose file''. This would at
least work for the cases where we would have used $C_1$ and $C_2$.
Yet, we still reduce the possibilities of $C_g$. Now, the dialog
can only be used when a file must be chosen. $C_g$ could in principle
ask for anything. Therefore, the element becomes more specific
when setting this information to a fixed value in the library.

One could argue, that in Fig.~\ref{fig:abstractions}, we could move the 
abstractness directly from the lowest to the highest level, i.e., go 
directly from $C_1$ and $C_2$ to $C$. Although this is what we often see in 
practice, in theory there is more to it. Implicitly, we always make a 
generalising abstraction before we use simplifying abstraction. We need to 
identify the parameters and their types that distinguish the less abstract 
artefacts before we unify them to a more abstract artefact. Hence, these 
two abstraction are needed and sufficient to describe the effects of 
abstraction in design.

\subsection{Discussion} 

Tab.~\ref{tab:influences} summarises which consequences generalising and 
simplifying abstractions have for the specificity and both types of 
complexity. Both kinds of abstraction obviously increase the abstractness. 
However, the main aim of generalising abstraction is to make the artefacts 
more generic. Hence, the specificity in relation to the artefacts on the 
lower level decreases but at the cost of higher dynamic complexity. To 
comprehend the consequences for the detail complexity, one needs to 
understand the relationships and effects of the parameters in the more 
abstract artefact. Whether the detail complexity is really reduced depends 
on the number of parameters that need to be introduced and the number of 
elements that are abstracted from. If abstracting away two elements requires 
the introduction of two parameters, the detail complexity stays the same. If, 
however, the number of parameters is smaller than the number elements 
abstracted away, the detail complex decreases.

\begin{table}[h]
\caption{The influences of abstraction on specificity
         and complexity}
\label{tab:influences}
\begin{center}
\begin{tabular}{l|ccc}
\hline
Type & Specificity & Detail com. & Dynamic com.\\
\hline
Generalising & - & 0/- & +\\
Simplifying  & + & - & +/-\\
\hline
\end{tabular}
\end{center}
\end{table}

Simplifying abstraction reduces the detail complexity but with the drawback 
of higher specificity. By removing the parameters from the artefact, we 
obviously reduce the number of units and thereby reduce detail complexity. 
However, it might be that we increase dynamic complexity because while 
using the artefact we cannot control it by the parameters. Hence, the 
cause-effect structure might not be easy to understand. These basic rules 
are often not considered during software design. If we want to reduce 
complexity without increasing specificity, we can only reduce the detail 
complexity, often with the drawback of increasing dynamic complexity. A 
reduction of dynamic complexity will always increase specificity.

\section{Examples}
\label{sec:examples}

To foster the understanding of the concepts presented in this paper, this 
section presents a number of examples from different domains and 
exemplifies how our considerations apply to them.

\subsection{Natural Languages}

Natural languages like software systems undergo a continuous evolution. A 
central part of this evolution is the extension of languages through the 
establishment of abstractions. As in software development the goal of this 
abstractions is usually improved reuse and reduced complexity.

A recent example is the word ``Blog''. Before the concept of a Blog was 
known, people simply referred to the specific website they were talking 
about. As these website shared a number of characteristics like being 
updated frequently or being similar to an online journal the concept of 
``Blog'' emerged. This abstraction allowed people to communicate about a 
whole class of similar websites. However, at that time the concept was not 
lexicalised and one had to use lengthy circumlocutions to refer to it. 
Obviously, the main motivation for introducing a single term describing 
the concept was reuse. It spared people from using lengthy descriptions 
again and again. 

A Google search for \texttt{define:blog} yields about two dozen definitions 
of the term ``Blog'' that agree to a great degree on what a Blog is. 
Nevertheless, some definitions contain information others lack, \eg that a 
Blog usually reflects the personal opinions of the author or that Blog 
entries are typically small. 
According to our analogy these details
can be viewed as the \emph{parameters} of the concept ``Blog''. This shows
that the generalising abstraction does indeed increase the complexity.
Before the concept ``Blog'' was introduced one could simply refer to
website $X$ or website $Y$. Using the abstracted concept ``Blog'' he
needs not only to point that something is a Blog but also specify the
\emph{parameters} of this Blog.

As natural languages also aim at simplification the explicit specification 
of these parameters is usually omitted. Simplifying abstraction is used to 
abstract from the parameterised concept ``Blog'' to a parameterless concept 
with the same name. This example shows that abstraction also means loosing 
information: The parameterless concept ``Blog'' omits a number of details 
one needs to know to properly use it. As users of natural languages
are intelligent human beings (as opposed to software compilers) they are usually
capable of reconstructing this lost information from the context and 
their common knowledge. In certain situations, however, this information
loss can pose severe difficulties for human communication, too.

\subsection{Libraries}
The development of libraries is a showpiece of the abstraction mechanisms
described in this paper as their central goal is to abstract from
complex underlying functionality and present it to their users in an
easy-to-use manner. Examples for libraries can be found at several 
complexity levels and for various domains. Very well-known instances are
\emph{Java Swing} for building graphical user interfaces, 
\emph{Log4X} to support logging functionality and 
\emph{Java Collections}/\emph{C++ STL} that provide commonly used data
types. 

A good example for the discussed types of abstractions is the GUI library 
\emph{Java Swing} that allows developers to implement the elements and
concepts typically found in user interfaces, \eg windows, button, menus or
listeners. In comparison to implementing a window in Java
without such a library it doubtlessly raises the abstraction level and
reduces complexity. In fact, this and similar libraries are so widely
used that most developers do not know most of the details involved in
drawing a window on the screen, making it aware to mouse events, keyboard
events, etc. The library achieves this by presenting a well-chosen
simplifying abstraction of the underlying details and thereby enables even
casual users to quickly implement a program that opens a simple window.

However, the library can appear strikingly complex if one wants to fine-tune 
certain specifics of its behaviour. Based on our considerations about 
abstraction mechanisms, we claim that this can only partially be blamed on 
bad library design but is mainly rooted in the fundamentals of the 
generalising abstraction. The process of \eg drawing a window is itself a
complex task with countless variation points, like \eg window size, colour, 
shape, resizeability, etc. As \emph{Swing} is designed to enable the user 
to control a good part of these variation points it has to make them 
explicit as parameters that do increase its complexity (\eg the 
\texttt{JFrame} class has well above 50 accessible parameters). 

Although \emph{Swing} is quite good at hiding this complexity by providing
further simplifying abstractions (\eg one needs to set only one parameter
on class \texttt{JFrame} to show an empty window with a title) its
usability suffers from the mere number of language elements it contains.
Following our considerations in Sec.~\ref{sec:abstractions} we believe
that this cannot be improved without limiting its genericity and thereby
consequently narrowing the options a user has to fine-tune the user interface
he implements.

\subsection{Domain-Specific Languages}
Domain-specific languages (DSLs) are built with the purpose to abstract
from the solution language and enable their users to interact with concepts
close to the problem domain. Prominent examples are \emph{SQL} for
database access, \emph{BNF} for syntax specification, \emph{ANT} for
building software, \LaTeX\ for type setting and the \emph{DOT} language
used by the graph layout tool \emph{GraphViz}%
\footnote{\url{http://www.graphviz.org/}} to specify graphs.

The \emph{DOT} language provides a good example for the discussed types of 
abstractions. Together with the \emph{GraphViz} tool it allows the creation of 
directed graphs with highly sophisticated layouts without knowing the least 
bit about graph layout algorithms. Due to its use of simplifying 
abstraction the most simple graph can be specified by the single line: 
\texttt{A~->~B}. Although the language allows some control of graph attributes 
like colours, fonts, line width, etc. it offers only limited options to 
influence the way graphs are laid out. While this may be perceived as
limitation in specific situations, the achieved reduction of complexity
enables users to efficiently solve their graph drawing problems in most
situations.

One notable property that the \emph{DOT} language shares with many other DSLs
is its lack of mechanisms for defining own abstractions. In contrast
to libraries that support new abstractions by using the mechanisms of
their host language, DSLs often lack such mechanisms as this
would require the language designers to explicitly create them.

\subsection{Programming Languages}

When writing programs with general purpose programming languages like C or 
Java, developers continuously build or extend languages by introducing new 
abstractions~\cite{2006_ratiud_programs_knowledge_bases}. 
Even the definition of a constant like 
\texttt{VALUE\_ADDED\_TAX\_RATE = 0.16} is in fact an abstraction as it
abstracts from a concrete number and introduces a new language element.

For more complex concepts the abstractions are typically realised by defining 
first-class program elements. An example is 
the search for a specific character in a string. This can be captured
in a function to foster reuse.
In its most concrete form 
such a function searches for a specific character in a 
specific string using a specific search algorithm. 
As this function is not reusable the
developer has to abstract from the specifics to implement a function
that can be used in multiple contexts. Therefore, he has to create a function
with a number of parameters. 
These parameters include the character to search for and the string
to be searched. It abstracts from a specific character and a specific
string to all characters and all strings. 
The sort algorithm is parameterised to distinguish
between case-sensitive and case-insensitive search. 

As even this seemingly simple function is equipped with up to three 
parameters, this example shows how generalising abstraction increases 
detail complexity. Later the developer may find this function too complex 
and unhandy to use. He could then introduce a simplifying abstraction by 
introducing a new method that always performs a case-insensitive search. By 
doing so he reduces detail complexity and genericity. However, dynamic 
complexity can be increased because the user of the function needs to be
aware of these options.

Looking at the whole language defined by the program, the question how this 
last step affected the abstractness, genericity and detail complexity of this 
language, is unfortunately not straight-forward to answer. An important 
factor is the fate of the original search function. If it was 
removed the language's abstractness was increased while its genericity and 
complexity were reduced. If the new function was included in addition to 
the old function the the situation is more complex.
The 
genericity of the languages did not change as the language still provides 
access to the old function. The complexity of the language, however, is a 
matter of discussion: On one hand one could say the language became more 
complex as it grew by one element. On the other hand one could claim that
the language became less complex as it offers a simpler entry point now
(the new search function). We believe that this cannot
be answered in general and depends on the context as well as on the
language design goals.

\subsection{Modelling Languages}

There is obviously no clear distinction between modelling languages,
domain-specific languages and programming languages. However, modelling
languages are typically considered to have a higher level of abstraction as
programming languages and they often come with a graphical notation to
support comprehension. Well known examples are the Unified Modelling Language
(UML), Matlab Simulink and Stateflow, or Entity-Relationship diagrams (ER).

Recently, the model-driven architecture
(MDA)\footnote{\url{http://www.omg.org/mda/}} approach has received
considerable attention in research and practice. Its aim is to use simple UML
models that are transformed several times down to running code. The promise
is that the productivity in building the simple -- abstract -- models is much
higher. This approach is a perfect example for the relationships described in
Sec.~\ref{sec:abstractions}. In current MDA tools the models that are mainly
used are UML class diagrams. These diagrams cannot express a lot of detail
about behaviour. The emphasis is on the data and structure similar to an
entity-relationship diagram. In order to generate a runnable program from
this model a great deal of additional information needs to be added. Hence,
the fixed (and hidden) part is extensive. It depends largely on the tool and
the way the transformations are described. This large amount of fixed
information implies that the context in which the model can be used, i.e.~the
type of software that is generated from it, needs to be rather specific.
Today, several MDA tools are able to generate simple, web-based database
applications. However, if we needed to generate a different application, the
abstraction provided would be useless.

Another example widely used in the embedded systems domain is the
Simulink and Stateflow toolkit provided by the Matlab tool. Simulink is a
graphical dataflow language that provides blocks for various continuous and
discrete functions. They can be combined to calculate the results needed.
Stateflow is a statecharts dialect that can be used in combination with
Simulink blocks in order to model the more state-oriented parts of the
application. There are some code generators available for these modelling
tools that typically generate C code.

The Simulink/Stateflow toolkit is a prominent example for successful 
abstraction. The toolkit offers the abstractions typically needed for 
embedded systems design, parts of differential equations as well as state 
transitions. This frees the designer from a lot of the ugly details that 
are involved in embedded systems, especially the concrete hardware details 
and the interfaces to the platform software. However, for this to be 
possible, the toolkit is rather specific. It is useful for the embedded 
domain but to use it in web development is rather impossible because the 
data structure parts are not strongly supported. The abstractions 
chosen simply are not able to help in data design. We have again a certain 
amount of fixed information in the code generators. This information 
determines the contexts in which the language can be used and hence its 
specificity.

\section{Consequences}
\label{sec:consequences}

We identified
two main types of abstraction in software design
and showed how they affect the three characteristics abstractness, specificity,
and complexity of the designed artefacts. These definitions were 
substantiated by several examples and
a survey. Based on this, we are able to describe several consequences
and insights that follow from it.

\paragraph*{Generalising Abstraction Increases Dy\-na\-mic Complexity} 
A goal for language
development is often to increase the abstractness in order to reduce
complexity but also to decrease or maintain the
specificity, e.g., \cite{bernholdt04}. We showed that generalising
abstraction is able to increase abstractness by increasing the
genericity. However, this is only possible by increasing the dynamic
complexity which is in fact contrary to the initial goal of
complexity reduction. The consequence is that we cannot increase
genericity and reduce dynamic complexity at the same time. We are only able
to reduce detail complexity. Hence, we need to decide on the trade-off
between detail and dynamic complexity with respect to the expected reuse and
understandability issues.

\paragraph*{Simplifying Abstraction Increases Specificity} Abstraction is often
proposed as the key to reduce complexity in software development. We
showed that this is only true for simplifying abstraction. We loose
information about some parts of the artefact.
Hence, it is less complex and thereby becomes more comprehensible.
However, we also saw that simplifying abstraction always implies a
loss of genericity, i.e., an increase in specificity. Therefore,
when we increase abstractness and reduce complexity, we must be
aware that we need to increase the specificity. This applies, for
example, to the model-driven architecture approach (MDA).
This approach
uses abstract models that -- if not excessively parameterised --
can only generate software for a specific, predefined domain.

\paragraph*{To Manage Complexity, Design Specific and Generic} In general,
it makes sense to have abstract, specific artefacts that are useful
for new developers and common, often occurring situations. It
is helpful for expert developers to have also more generic
elements so that they can express their seldom occurring problems. This
avoids detail complexity to a certain extent and hence supports reuse as well
as understandability of the design.

\section{Related Work}
\label{sec:related}

There is a surprisingly small amount of discussions of abstraction and its 
relationships with complexity and specificity (or generality) in the 
literature on software design. One exception is Kramer's contribution in 
~\cite{2007_kramerj_abstraction}. He also states that there are two types 
of abstractions, a simplifying and generalising one. However, the article 
focuses on the abstraction skills of computer sciences students and, hence, 
does not discuss the consequences of this definition for software design.

Obviously, the abstractions introduced and used in common programming
languages have been discussed, e.g.\
\cite{shaw84}.
An area in software design where abstractions are used explicitly is 
design patterns \cite{gamma95}. However, also the pattern community fails to
look at the effects on other properties of the design artefacts.

Krueger describes in \cite{krueger92} ideas that we used as a
basis for several aspects of this paper. He states that abstraction 
is an essential
part of reuse and that this has been noted by several other authors.
It helps developers in selecting and specialising
artefacts.
He also remarks
that a ``generalized reusable artifact is in fact an abstraction
with a variable part.'' Finally, he uses the concept of \emph{cognitive
distance} that is similar to what we call complexity. In contrast to this
work, we contribute the clear distinction between the two types of
abstraction and their effect on the abstractness, specificity, and
complexity of a design.

Because we consider also the elements of a (domain-spe\-ci\-fic)
language as artefacts, the general literature on the
design of programming language is also relevant
\cite{appleby91,friedman92,ghezzi82,louden02}. It defines several properties 
or characteristics of languages. The \emph{readability} is a desired
property of a language. This is related to abstractness and
complexity. A language that is simple and uses abstract concepts
is more readable than a language with many elements and technical
details. The complexity of a language is also recognised as an
important principle in language design by requiring \emph{simplicity}.
However, a definition of simplicity and its tradeoffs is not given
in the literature. Related to our work is also the principle of
\emph{programming efficiency} or \emph{expressiveness} of programming
languages. It describes the easiness to express complex processes
and structures. We can describe this issue by the abstractness and
the specificity. Very abstract language elements (using simplifying
abstraction) allow to express those complex processes and structures
concisely. The genericity of such an element then allows to alter
certain parts only. Hence, this also contributes to programming
efficiency.

Our work is also related to the work on domain-specific
languages, e.g., \cite{mernik05}. It is explained there that
the design of a domain-specific language must pay off in terms
of more efficient development and maintenance. However, the essential
tradeoff between specificity and complexity is not explicitly stated.
It is also noted that a
domain-specific language can be an application library or simply
embedded into a so-called general purpose language by abstract data
types. This supports
our view that all kinds of software design are similar to this respect. 

Prenninger and Pretschner discuss abstractions for model-based
testing in \cite{prenninger04}. Although this is not general software
design the essential ideas apply. They also see abstraction
as losing information that can either be automatically inserted
or not. The main goal for abstraction is given as simplification
because generalising abstraction is not discussed in that paper.
However, it is also stressed that abstraction (especially the
automatically resolvable one) tends to be highly domain-specific.
We are able to show why this is the case using our abstraction types.

Finally, the paper of Bernholdt, Nieplocha, and Sadayappan \cite{bernholdt04} 
is an example that neglects the tradeoffs discussed in this paper.
They classify languages for high-performance
computing in the dimensions ``abstraction'' and ``generality''. Based
on this it is
stated that the ultimate goal of further language developments should
be to rank high in both dimensions, i.e., that new languages should be
more abstract and more general at the same time. We show that this is
only comes at the cost of higher dynamic complexity although their first goal
was complexity reduction.

\section{Conclusions}
\label{sec:conclusions}

Abstraction is an essential activity in software design. We use it as a 
means to improve understandability by reducing complexity and to support 
reuse of the artefacts developed. However, the effects of the abstraction 
and hence the abstractness of artefacts is rarely discussed and not well 
understood. The extremes -- very concrete or very abstract -- are clearly 
not the aim in software design. There must be a reasonable choice somewhere 
in between. For this choice to be made, we need to know the effects and 
trade-offs involved.

For this, we introduce three characteristics of design artefacts: 
abstractness, specificity, and complexity. We show that there are two basic 
types of abstractions that have different influences on these 
characteristics. \emph{Generalising abstraction} decreases the specificity 
and detail complexity of artefacts but also increases the dynamic 
complexity, \emph{simplifying abstraction} decreases again the detail 
complexity but at the cost of higher specificity. The simplification can 
even increase the dynamic complexity. Obviously, both abstractions increase 
the abstractness. Based on these insights and several examples, we are able 
to formulate several consequences on software design.

We are fully aware that so far this is a rather theoretical work. It is 
quite probable that we will never be able to show these relationships 
empirically because there is a certain amount of subjectivity involved. For 
the detail complexity, we might be able to count the details and compare 
complexity but cause-effect relationships are not that easy to measure. 
Moreover, the specificity of an artefact is probably not measurable. How 
should we count the number of contexts where it can be used? Nevertheless, 
we see several relationships that hold in this fuzzy environment that in 
our view are useful to understand and to consider in software design in 
order to build good abstractions.

For future work, we plan to concretise the ideas presented here by building 
an \emph{abstractness-complexity calculus} that allows us to express the 
interrelations in a more formal way. In general we aim at applying the 
developed concepts to different areas of software engineering. We see 
strong relations to the areas of aspect-oriented programming as well as 
product-line engineering. In these areas, our approach might help to 
explain the concepts and especially the shortcomings.

\subsection*{Acknowledgements}
We are grateful to Manfred Broy, Martin Feilkas, Elmar J\"urgens and 
Daniel Ratiu for fruitful discussions on the topic.
\balance

\bibliographystyle{abbrv}
\bibliography{rase08}

\end{document}